\documentclass[prb,aps,twocolumn,showpacs]{revtex4}
\usepackage{graphicx}
\usepackage{dcolumn}
\usepackage{amssymb}
\usepackage{amsmath}

\begin{document}

\title{Three-dimensional molecular dynamics simulations of void
coalescence during dynamic fracture of ductile metals }

\author{E.\ T.\ Sepp\"al\"a}\thanks{Present address: Nokia Research Centre, 
P.O.\ Box 407, FIN-00045 NOKIA GROUP, Finland}\email{eira@iki.fi}
\author{J.\ Belak}\email{belak@llnl.gov}
\author{R.\ E.\ Rudd}\thanks{Corresponding author}\email{robert.rudd@llnl.gov}

\affiliation{Lawrence Livermore National Laboratory, Condensed Matter Physics 
Division, L-045, Livermore, CA 94551, USA}
\date{\today}

\begin{abstract}
Void coalescence and interaction in dynamic fracture of ductile metals
have been investigated using three-dimensional strain-controlled
multi-million atom molecular dynamics simulations of copper. The
correlated growth of two voids during the coalescence process leading
to fracture is investigated, both in terms of its onset and the
ensuing dynamical interactions.  Void interactions are quantified
through the rate of reduction of the distance between the voids,
through the correlated directional growth of the voids, and through
correlated shape evolution of the voids.  The critical inter-void ligament
distance marking the onset of coalescence is shown to be approximately
one void radius based on the quantification measurements used,
independent of the initial separation distance between the voids and
the strain-rate of the expansion of the system. 
The interaction of the voids is not reflected in the volumetric 
asymptotic growth rate of the voids, as demonstrated here. Finally,
the practice of using a single void and periodic boundary conditions
to study coalescence is examined critically and shown to produce
results markedly different than the coalescence of a pair of
isolated voids. 
\end{abstract}


\maketitle

\section{Introduction}
\label{sec_intro}

The fracture of ductile metals at high strain rates has been
understood at the microscopic level as a process of nucleation, growth
and coalescence of voids.~\cite{DynFract,Barbee} Initially voids
nucleate at the weak points in the material such as inclusions and/or
grain boundary junctions.  Once nucleated, the voids grow under the
tensile stress, driven by the reduction in elastic energy.
Eventually, the voids grow sufficiently large that they interact with
each other, in some cases link through localized shear, coalesce into
larger voids, and finally form the fracture
surface.~\cite{Curran,McClintock} Considerable experimental and
theoretical work has gone into the development of our understanding of
fracture. The fracture process has been modeled at various levels, but
most of the work involving the simulation of the activity of
individual voids has concentrated on the void growth and its
relationship to the plastic deformation of the surrounding material,
including effects such as the localization of this shear deformation.
Relatively little work has gone into the explicit modeling of void
coalescence, and both the understanding of the physics of this process
and the knowledge of how it should be implemented robustly in
continuum fracture codes remain open issues.  The point at which voids
begin to coalesce during dynamic fracture is of considerable interest
because complete fracture of the material typically ensues rapidly
thereafter.  As new experimental techniques have constrained the void
growth models ever more stringently,~\cite{belak4} a real need for a
well developed theory of coalescence has arisen.

Computationally void growth has been studied extensively at the
continuum level,~\cite{RiceTracey,McClintock2,Gurson,tvergaard} also
in dynamic, high strain-rate, conditions.~\cite{zikry,cortes} Recently
we have studied void growth at the atomistic level under high
strain-rate expansion, motivated by spallation
experiments.~\cite{belak,belak2,belak3,rudd,prb} The atomistic studies
demonstrate that voids grow by emitting dislocations that carry away
the material, platelets of atoms, from the void and are responsible
for the plastic deformations needed to accommodate significant void
growth.  There are also many recent studies of fracture in ductile
metals with several holes or
voids.~\cite{koss,tonks,pardoen,zohdi,benson} 
While these studies model the void
growth explicitly with fairly sophisticated models of plasticity in many
cases, they typically simplify the coalescence process to instantaneous
unification of the voids when some threshold is reached, such as once 
the voids grow to within one diameter of each other.
There are also
studies that investigate the competition between void by void growth
versus multiple void interaction in crack
propagation.~\cite{tvergaard2}  Several earlier continuum
studies~\cite{pardoen2,orsini,becker} and the one atomistic study
known to us~\cite{somerday} of the coalescence process have been
typically conducted in effectively two dimensional and highly
symmetric systems.

This Article covers in detail a study of the onset of void 
coalescence.  The first results of this study have been presented 
in a Letter.~\cite{prl_trial} 
Here we provide a more complete presentation of the results of
our study of void coalescence.  In addition to a more detailed 
description of the results presented in the Letter, we describe
different measures of void interactions such as shape changes induced
by a neighboring void, additional analysis of the behavior of the
system including stress-strain curves and void volume curves, and 
a new analysis of how the coalescence of isolated pairs of voids 
differs from that of voids in highly symmetric periodic arrays.

In particular the goal of this Article is to quantify the point 
at which coalescence begins, as measured by a
critical {\it inter-void ligament distance} (ILD), and examine the
mechanisms involved in the transition from independent void growth to
coalescence.  There are several ways in which two voids can interact.
In the case of pure impingement, the voids only interact when they
grow to the point that they intersect and join into a single void.  In
reality, the voids interact before they intersect.  Their range of
interaction is extended due to their elastic and plastic fields.  Each
void generates an elastic strain field of the form generally
associated with centers of dilatation.~\cite{Eshelby} The shear stress
decreases with the distance from the void like $r^{-3}$.  For voids
sufficiently close each void's growth rate is altered by the stress
field of the proximal void.  The modification of the elastic field can
affect the initiation of plasticity, as well as the subsequent
development of the plastic zone around the voids. The voids may
interact through their plastic fields, too, in which case the fields
may give rise to an increased hardening rate in a localized region or
to thermal softening and shear localization.  An argument due to Brown
and Embury for a transition to shear deformation based on simple
geometrical considerations suggests that the critical inter-void
ligament distance, ILD$_c$, should be equal to one diameter of a
void;~\cite{Brown} that is, when the surfaces of a pair of voids are
separated by one void diameter, they transition from independent void
growth to coalescence.  It is at this point, they argue, that the
dominant void process switches from the radial plastic flow around
isolated growing voids to a large-scale shear deformation allowing the rapid
coalescence of the pair of voids.  However, more recent
two-dimensional studies suggest that for distances between voids as
large as six diameters the void growth rate is
enhanced.~\cite{horstemeyer} 

The use of atomistic techniques permits the analysis of the
contributions of these competing mechanisms to the onset of void
coalescence, as we describe in this Article. We demonstrate the
existence of, and compute, the critical inter-void ligament distance
ILD$_c$ by starting with two voids well separated from each other and
detecting the point at which correlated growth begins, marked both by
the accelerated rate at which the two void surfaces approach each
other and by biased growth causing the voids to start to extend toward
each other. These changes give an indication of the onset of the coalescence
process, and it tests the argument by Brown and Embury.~\cite{Brown}
We also test the setup by Horstemeyer {\it et al.}~\cite{horstemeyer}
by varying the initial distances between the voids and measuring the
asymptotic growth rate of the voids.  The initial void-to-void
distance below which the growth-rate is enhanced should give another
candidate for the critical distance and measure it in a volumetric
sense. It should be noted, however, that the three-dimensional 
void coalescence studied here with molecular dynamics, as indeed any
3D coalescence of roughly spherical voids, does not admit
the two-dimensional shear mechanism proposed by Brown and Embury 
in its simplest form, and therefore these different 
analyses are not fully comparable. 
Also Horstemeyer {\it et al.} used a more symmetric setup
than the simulations covered in this Article.

The Article is organized as follows. The simulation method and the
performed computations are introduced in Section~\ref{sec_method}.
The basic mechanism, dislocation driven void growth, is demonstrated
and the key reference parameter, mean linear void size, is introduced
in Section~\ref{sec_voidgrowth}. The interaction between voids are
studied in Section~\ref{sec_interaction}: Section~\ref{sec_dist} uses
two different distance measurements to study the interaction;
Section~\ref{sec_shape} introduces a shape parameter for the purpose.
Finally the volume effects of the void growth are studied in
Section~\ref{sec_volume}: two separate voids are studied in
Section~\ref{sec_volume_twovoids} and one void and the interaction
with its periodic image are studied in
Section~\ref{sec_volume_onevoids}. The paper ends with
conclusions, Section~\ref{concl}.

\section{Method and Simulations}
\label{sec_method}

\begin{figure}
\includegraphics[width=75mm]{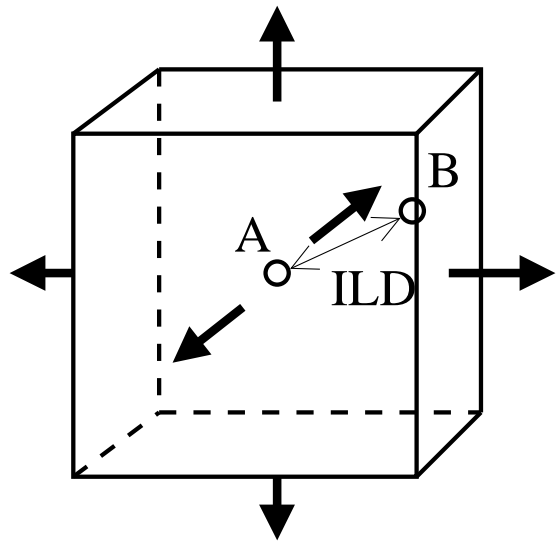}
\caption{A sketch of the simulation configuration. The simulation box 
includes $120\times120\times120$ FCC unit cells 
with periodic boundary conditions
for a total of 6~912~000 atoms initially.  From this two equal size 
voids are created by removing approximately 3620 atoms for each void. 
The thin arrow shows
the inter-void ligament distance (ILD), the shortest surface-to-surface
distance between the voids, identified as A and B.
The initial ILD ranges between one-half and five times the 
initial void radius.
The bold arrows denote
triaxial, hydrostatic, expansion of the box. The strain-controlled
expansion is applied with constant strain-rates of
$\dot{\varepsilon}=10^8$/sec and $10^9$/sec.}
\label{fig1}
\end{figure}

\begin{table}
\caption{\label{tab1} The initial position for void B relative to
void A and the center-to-center distances for various initial
inter-void (surface-to-surface) ligament distances ILD$_0$ used in
this study.}
\begin{ruledtabular}
\begin{tabular}{ccc}
ILD$_0$ & position of void B & separation between the\\
 & relative to void A & centers of voids A and B \\ \hline
0.50 & $[0.1334,0.0622,0.0290]L$ & $0.150 L = 6.510$ nm \\
1.00 & $[0.1778,0.0829,0.0387]L$ & $0.200 L = 8.680$ nm \\
1.20 & $[0.1956,0.0912,0.0426]L$ & $0.220 L = 9.548$ nm \\
1.50 & $[0.2223,0.1037,0.0483]L$ & $0.250 L = 10.85$ nm \\
1.81 & $[0.2500, 0.1166, 0.0544]L$ & $0.281 L = 12.20$ nm \\
4.62 & $[0.5000,0.2332,0.1087]L$ & $0.562 L = 24.40$ nm \\
\end{tabular}
\end{ruledtabular}
\end{table}

We have performed a series of large-scale classical molecular dynamics
(MD) simulations~\cite{allen} in single crystal face-centered cubic
(FCC) systems using an empirical embedded-atom model (EAM) potential
for copper.~\cite{oh1,oh2} The three dimensional
(3D) simulation box consists of $120\times120\times120$ 4-atom FCC cells with
periodic boundary conditions for a total of 6~912~000 atoms, in most cases.
In Section~\ref{sec_volume_onevoids} where results from smaller system
size simulations are presented, the details of the sizes are
introduced.  For the nearly 7 million atom simulations 
the MD code was parallelized
using spatial domain decomposition and run in a massively parallel
computer using from 64 to 256 processors.

In the simulations the system is initially equilibrated using a
thermostat~\cite{hoover} at room temperature, $T=300$~K, and a
constant volume $L^3$ (with $L=43.3~\mbox{nm}$) chosen to give ambient
pressure, $P \simeq 0$~MPa.  Once the system has reached equilibrium,
atoms are removed to create two spherical voids in the system with
radius $r_0 = 0.05L = 2.17~\mbox{nm}$: one centered in the box and the
other located a distance ILD$_0$+2$r_0$ away in the direction 
$\hat{u} = [0.8892054, 0.41464327, 0.19335135]$ from the first void.  
We refer to these as void A and void B, respectively, see Fig.~\ref{fig1}.  
ILD$_0$, the {\it initial
inter-void ligament distance}, is the closest surface-to-surface
distance between the voids~\cite{comment} and it is varied here, but
the relative orientation of the voids is kept fixed.  For ILD$_0$ we
have used the values: 1.00, 1.20, 1.50, 1.81, and in some cases 0.50
and 4.62. The unit for the ILD is the void diameter, $d = 2r$.  The
positions for the center of the void B and the distances between the
voids are listed in Table~\ref{tab1}. Initially, the voids are equal
in size, with approximately 3620 atoms removed for each. This removal
of atoms can be interpreted as an instantaneous debonding of two
infinitely weakly bound inclusions.

Once the voids are formed, the thermostat is turned off, and
dilatational strain is applied uniformly at a constant strain-rate
$\dot{\varepsilon}$.  The strain-controlled simulations~\cite{belak}
are carried out using the scaled coordinate formulation typically
employed in the constant pressure method due to
Parrinello and Rahman.~\cite{parrinello} Use of scaled coordinates
prevents the
spurious generation of elastic waves at the box boundaries. In this
method the positions of the atoms are stored using rescaled
coordinates between $[0,1)$. When calculating the forces and new
positions of the atoms their coordinates are multiplied by a diagonal
scaling matrix ${\mathcal H} = \{L_x,L_y,L_z\}$. This scaling matrix
is updated each time step, when the expansion is applied, by
multiplying the initial matrix ${\mathcal H}_0 = \{L,L,L\}$ with the
sum of the unit matrix and the strain matrix ${\mathcal E} =
t\dot{\mathcal E}$, ${\mathcal H}(t) = {\mathcal H}_0 ({\mathbb I} + t
\dot{\mathcal E})$. Applied strain-rates of
$\dot{\varepsilon}=10^8$/sec and $10^9$/sec have been used with
perfectly triaxial, or hydrostatic, expansion, $\dot{\mathcal E} =
\{\dot{\varepsilon}_x, \dot{\varepsilon}_y, \dot{\varepsilon}_z\} =
\{\dot{\varepsilon}, \dot{\varepsilon}, \dot{\varepsilon}\}$.  Thus
$L_x(t)=L_y(t)=L_z(t)=L(t) = V^{1/3}$, where $V$ is the volume of the
box.  A time step of 6.7 fs was used.
More details of the simulation method can be found in
Ref.~\onlinecite{prb}, including analysis of growth of a single void
of the same initial radius $r_0$ in non-triaxial expansion.

\section{Dislocations and Void Growth}
\label{sec_voidgrowth}

\begin{figure}
\includegraphics[width=75mm]{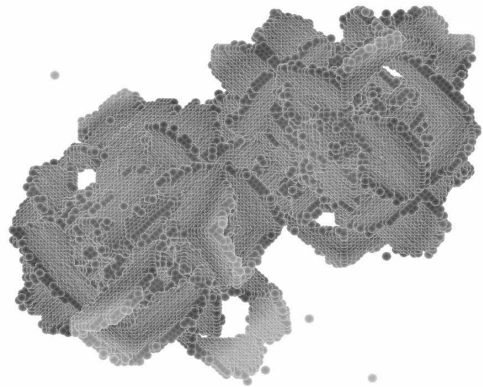}
\caption{(Gray-scale) A three dimensional snapshot of the two voids
with the prismatic dislocation loops forming from the voids.  Only
those atoms belonging to the void surfaces or to dislocation cores, stacking
faults or other defects are shown (and a small number of extraneous atoms
due to thermal fluctuations).  The stacking fault ribbons are the
broad plates between leading and trailing partial 
dislocations. The snapshot is from the simulation with the initial
inter-void ligament distance ILD$_0$=1.81 diameters, and the
strain-rate $\dot{\varepsilon}=10^9$/sec. 
The strain at the snapshot is $\varepsilon = 2.93 \%$.  }
\label{fig2}
\end{figure}

\begin{figure*}
\includegraphics[width=160mm]{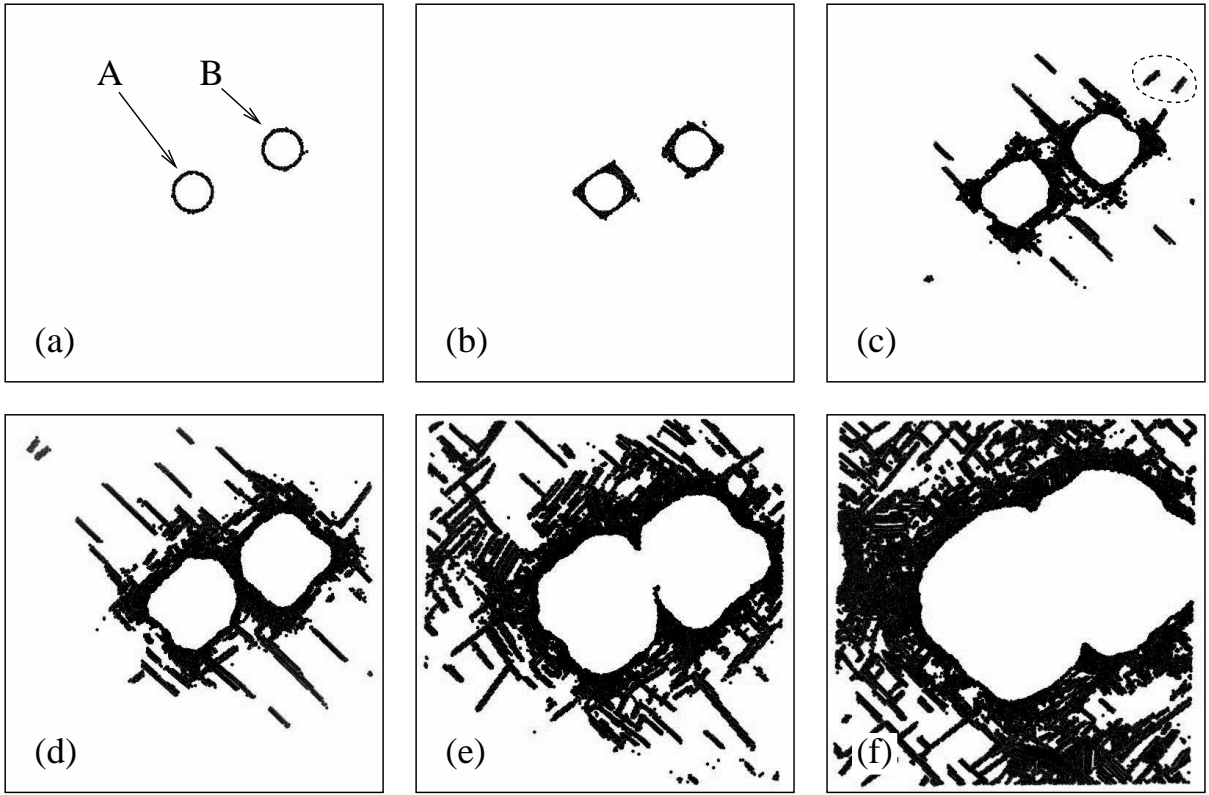}
\caption{Dislocation activity at six instants of time shown on one
particular slice through the system in order to expose the plasticity
near the void surfaces.  These snapshots are from the simulation with
$\dot{\varepsilon} = 10^9$/sec and again only those atoms in
dislocation cores, stacking faults, void surfaces or other defects are
shown (see text for more details and Fig.~\ref{fig2} for a full
three-dimensional figure). The dashed loop in panel (c) is drawn
around a slice of a prismatic dislocation loop.  The plane shown
passes through the centers of both voids with normal [0.145, 0.145,
-0.979]. The snapshots show the initial plasticity (a),(b), interacting
plastic zones (c),(d) and the final coalescence (e),(f).  The frames
correspond to the following values of strain $\varepsilon$: 
$1.72\%$, $2.42\%$, $3.47\%$, $3.89\%$, $4.52\%$, and 
$5.21\%$, respectively. }
\label{fig3}
\end{figure*}

\begin{figure}
\includegraphics[height=8cm,angle=-90]{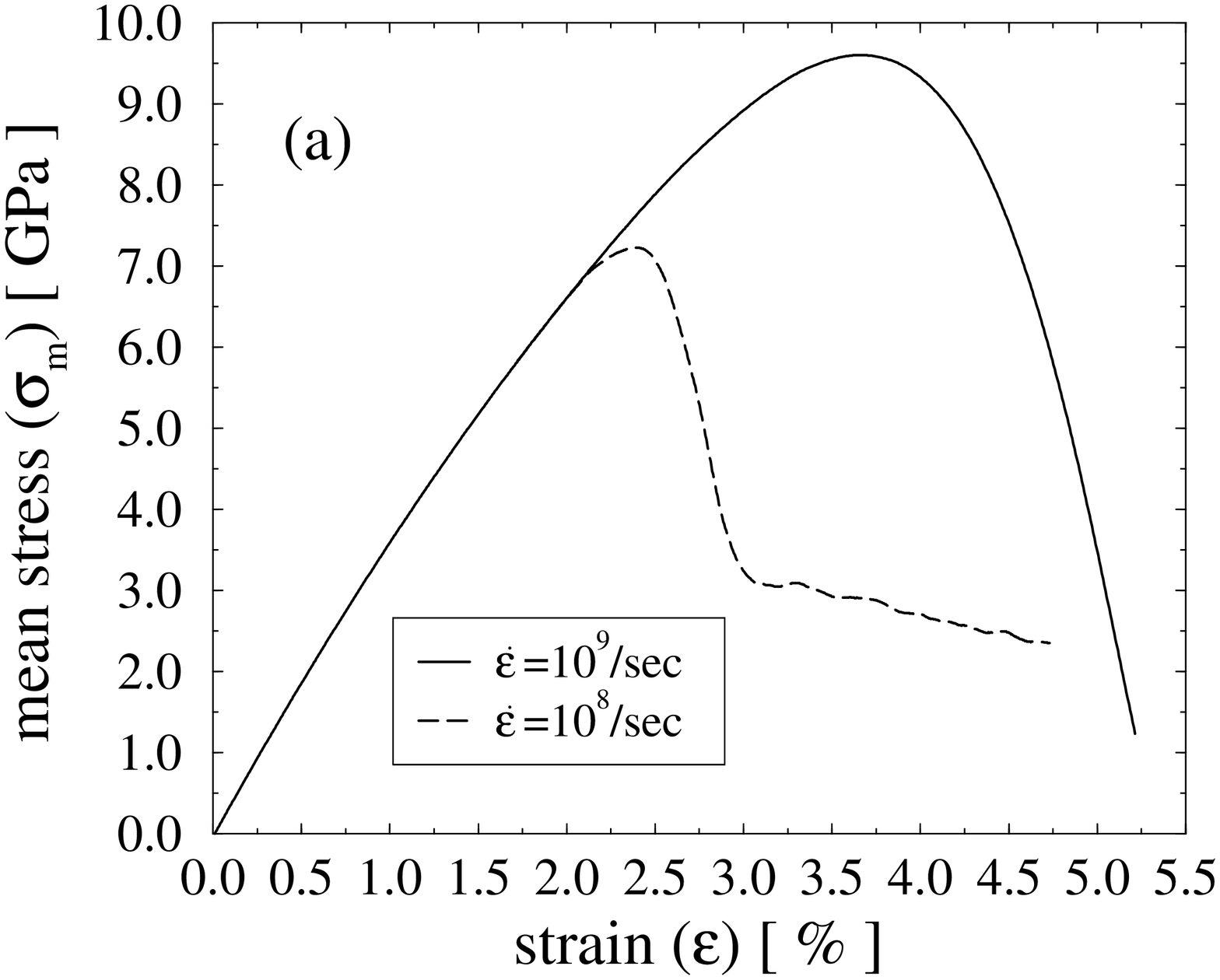}
\includegraphics[height=8cm,angle=-90]{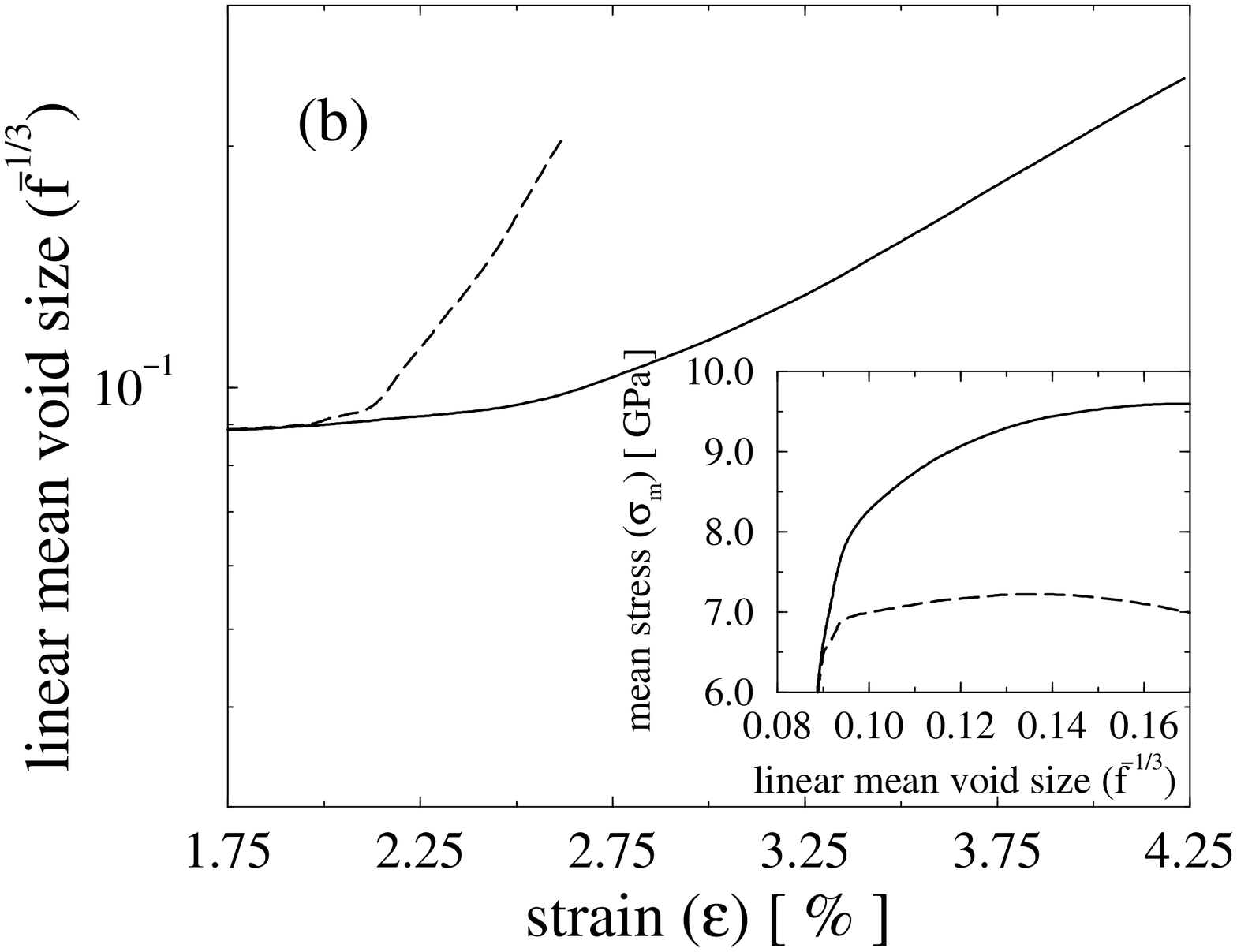}
\caption{(a) Mean stress $\sigma_m$ versus strain $\varepsilon$
from the simulations with the initial inter-void ligament distance
ILD$_0$=1.81 and the strain-rates $\dot{\varepsilon}=10^8$/sec (dashed
line) and $10^9$/sec (solid line). 
(b) Linear mean
void size, $\bar{f}^{1/3}$, (see the text for its definition) versus
the strain from the same simulations as the data in (a). The void
sizes are calculated until the coalescence of the voids.  The inset
shows the stress from (a) versus linear mean void size from (b).
Note the strain scales are different.}
\label{fig4}
\end{figure}

Let us start reviewing the simulation results by looking at some
figures to visualize growth of the  voids.  While some void growth
takes place through elastic stretching in the initial phases of the
box expansion, significant void growth and void-void interaction take
place only once plastic deformation has begun.  The important role of
plasticity leads us to consider in some detail how dislocations are
generated and the effect of dislocation dynamics on void
coalescence, which are the topics of this Section.

In Fig.~\ref{fig2} a three-dimensional snapshot of the voids is shown
from a simulation with the initial inter-void ligament distance
ILD$_0$=1.81 diameters and the strain-rate 
$\dot{\varepsilon}=10^9$/sec. In the plot only the atoms at
crystallographic defects such as void surfaces, dislocation cores,
and stacking faults are
shown. The decision of which atoms to plot is based on a geometrical
criterion, a finite-temperature generalization of the centrosymmetry
deviation.~\cite{rudd,hamilton} The power of this method of selecting
atoms based on the centrosymmetry deviation is that the visualization
can be done at finite temperature and on-the-fly, so that the system
need not to be cooled to zero-temperature and the atoms to be selected
based on potential energy, where quenching can influence the system
and prevent seeing the real configuration.  The snapshot in
Fig.~\ref{fig2} is when the voids have started to grow by emitting
dislocations, and thus the system has already evolved somewhat 
through plastic flow.  The broad plates in the figure are stacking fault
ribbons between leading and trailing partial dislocations. The
first generation of dislocation loops have not yet totally formed and
separated from the voids in this snapshot, but are about to do so, 
as can be seen as
their first halves have already separated from the void surface.

Figure~\ref{fig3} shows a series of visualizations of the crystal 
defects within a slice of width 4.5~{\AA} about a plane
including centers of both voids at six different instants during
coalescence. These snapshots are from the same simulation as in
Fig.~\ref{fig2}, and the same criterion for showing the atoms is used.
Now only a slice is shown in order to reveal the dislocation activity
near the void surface as the dislocation density increases. 
Figure~\ref{fig2} is a snapshot from a state of the system between
panels Fig.~\ref{fig3}(b) and (c).  From the snapshots it is apparent that
the deformation mechanism involves the nucleation and propagation of
dislocations, accommodating the void growth, and the interaction of
the dislocations.~\cite{HirthLothe}  
For example, the prismatic dislocation loops punched out by the voids
appear as roughly parallel line traces (due to the stacking fault
ribbons) in the slice Fig.~\ref{fig3}(c), as verified in the full 3D
configuration (cf.\ Fig.~\ref{fig2}).  Initially the
dislocation activity around each void is essentially symmetric
[Figs.~\ref{fig3}(a) and (b)], as expected for independent void growth,
but as the plastic fields evolve the void-void interaction is clearly
evident both through interactions between the two plastic zones and
bias due to the elastic fields [Fig.~\ref{fig3}(c)].  Once the
dislocation density grows sufficiently high in the ligament region
between the voids [Fig.~\ref{fig3}(d)], void B begins to grow in the
direction away from void A.  Next the voids coalesce
[Fig.~\ref{fig3}(e)], and continue to grow as one until ultimately
the void coalesces with the its periodic images [subsequent to
Fig.~\ref{fig3}(f)] so that the cavity percolates through the periodic
system.  The details of this final stage depend strongly on the
periodic boundary conditions and will not be of interest here.

The dislocation formation is closely related to the volume evolution of the
voids.  The voids grow by emitting dislocation loops, driven by the
reduction in the elastic energy as the increase in void volume allows
the strained matrix material to relax.  This relaxation
can be detected from the saturation of the increasing
stress in the system.  All of these phenomena happen simultaneously,
see Ref.~\onlinecite{prb}. In Fig.~\ref{fig4}(a) the mean
(hydrostatic) stress $\sigma_m$ is plotted with respect to
strain $\varepsilon$ (the control parameter in these simulations) for
strain-rates $\dot{\varepsilon}=10^8$/sec and $10^9$/sec with
ILD$_0$=1.81.  The stress is calculated with
the virial expression $\sigma_m = - \frac{1}{3V} \left(
\sum_i |\vec{p}_i|^2/ m_i + \sum_{i,j|j>i} \vec{r}_{ij} \cdot \vec{f}_{ij}
\right)$, where for atoms $i$ and $j$ the $\vec{r}_{ij}$ is the
relative position, $\vec{f}_{ij}$ is
the force, $\vec{p}_i$ is the the momentum and $m_i$ is the mass
(for a recent discussion of atomistic stress calculations see 
Ref.~\onlinecite{ZimmerStress}).
Comparing Fig.~\ref{fig4}(a) with Fig.~\ref{fig4}(b) it is
apparent that the start of the accelerated void growth 
due to plasticity is accompanied by, and indeed causes, 
the stress to plateau (at the same strain value) as the
elastic dilatation is relieved.
The void growth is shown in Fig.~\ref{fig4}(b) by plotting
linear void size $f^{1/3}$, where the single void fraction is
$f=V_{void}/V$ and $V$ is the instantaneous volume of the box 
at time $t$.
The technique for calculating the void volume $V_{void}$ is described
in Ref.~\onlinecite{prb}. The void growth, stress saturation
and even the void coalescence take place at
significantly smaller strains for slower strain-rates,~\cite{prb} as
can be seen from the figures, too.  Therefore we conclude that a
natural way to plot quantities from different strain-rate simulations
in the same figure is to plot them versus linear void size, $f^{1/3}$.
Plotting versus $f^{1/3}$ (a derived quantity) is preferred to
plotting versus the strain because it reduces strain-rate effects.
Thus by choosing $f^{1/3}$ from Fig.~\ref{fig4}(b) as the reference
quantity, the start of the deviation from the elastic behavior can be
synchronized for the different strain-rates, see an example for the
mean stress from the inset of Fig.~\ref{fig4}(b), where the mean
stress starts to saturate for both of the strain-rates at
$\bar{f}^{1/3} \simeq 0.09$. In the two-void case we have chosen to
use the mean void fraction $\bar{f}$, which is calculated as the
average void fraction of the two voids. 

Throughout the remainder of the
Article we will typically use the linear mean void size $\bar{f}^{1/3}$ 
as the reference quantity. The initial linear mean void fraction at ambient
pressure in these simulations is $\bar{f}_0^{1/3} \simeq
(6.0\times10^{-4})^{1/3} = 0.084$. The initial mean void size
$\bar{f}=6.0\times10^{-4}$ is somewhat larger than the value one gets
from $\frac{4}{3} \pi (r_0/L)^3 = 5.2\times10^{-4}$. This potential
source of confusion arises for
two reasons: first, atoms that have their centers within the radius
$r_0$ from the void center are removed for creating the void. On the
other hand, when the void volume is calculated the surface of the void
is defined based on the centers of the {\it remaining} surface
atoms. Second, the void surface relaxes somewhat after the void is
formed. For more discussion of the void volume calculation, see
Ref.~\onlinecite{prb}. For reference, the mean linear void size 
in Fig.\ \ref{fig2} is $\bar{f}^{1/3}=0.111$, and in the first
four snapshots of Fig.\ \ref{fig3} 
$\bar{f}^{1/3}=0.089$, $0.094$, $0.149$, and $0.195$, 
respectively.  After coalescence $\bar{f}$ is not measured.

\section{The Influence of the Neighboring Void}
\label{sec_interaction}

\subsection{Distance Measurements of the Voids}
\label{sec_dist}

\begin{figure}
\includegraphics[height=8cm,angle=-90]{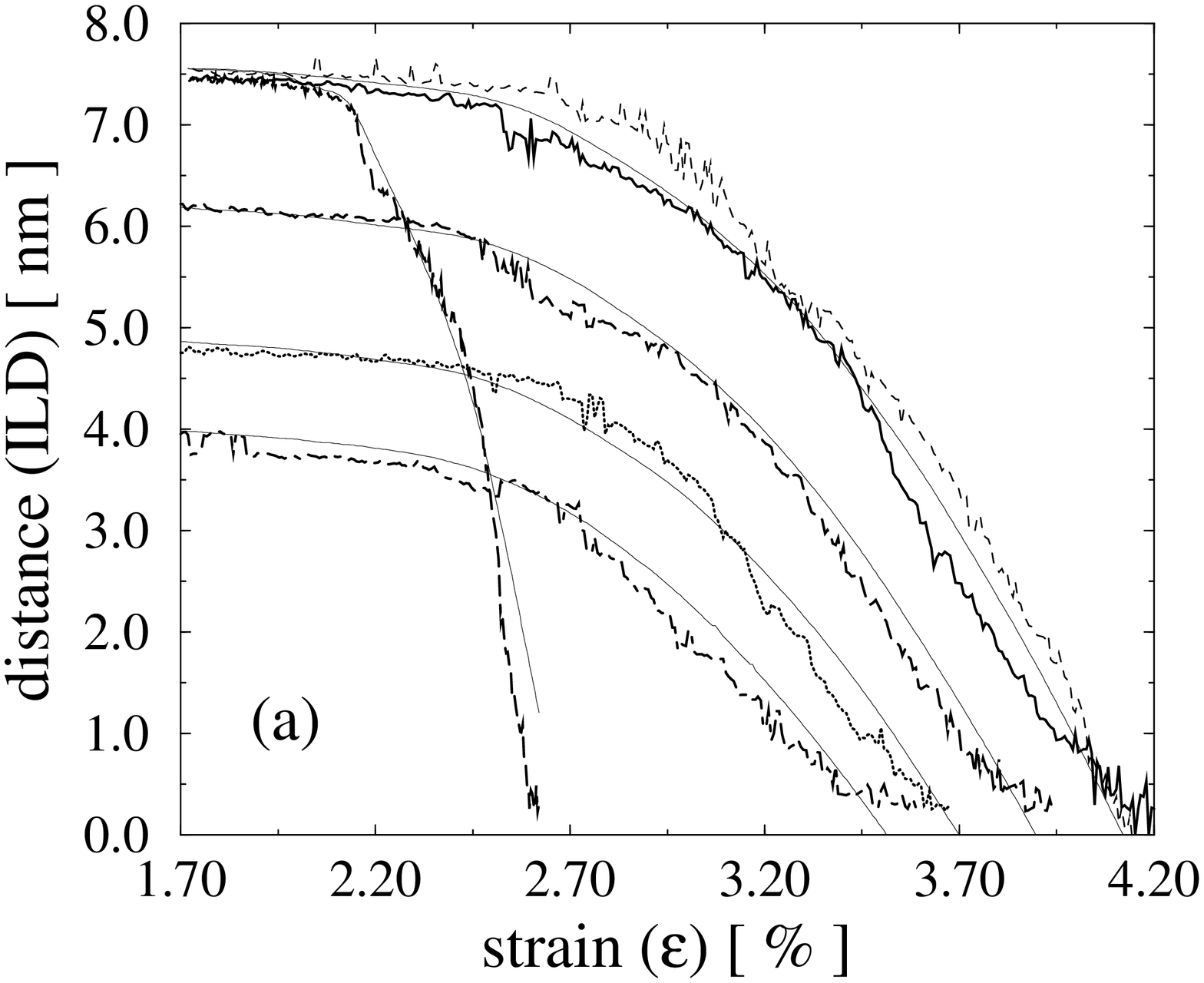}
\includegraphics[height=8cm,angle=-90]{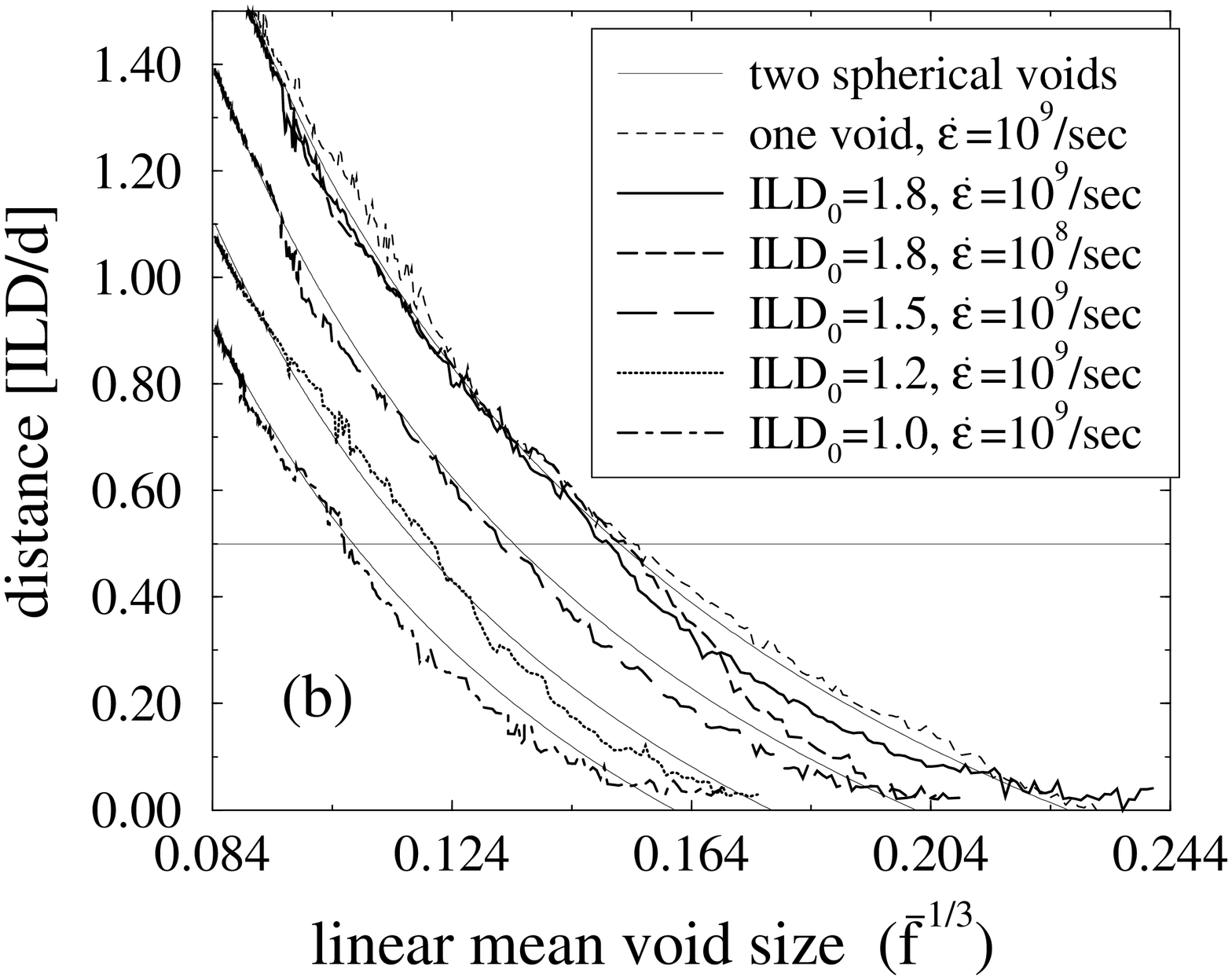}
\caption{Evolution of the ILD and the critical ILD. 
(a) Dynamical ILD, the distance between the surfaces of the
voids along the line connecting the original center positions for
various initial ILD$_0=$ 1.81, 1.50, 1.20, 1.00, plotted versus
strain.  
For ILD$_0$=1.81 the thick solid line and thick
medium dashed line denote $\dot{\varepsilon}=10^9$/sec and $10^8$/sec,
respectively. The thin solid lines show the hypothetical ILD for
spherical voids with the same ILD$_0$ impinging freely on each
other (see text). The short dashed
line shows the hypothetical ILD computed by duplicating a single void
(in the same box size as the two void simulations) at fixed centers
(here the duplication is to the position with ILD$_0$=1.81).  (b) The
same as in (a), but now plotted versus the linear average void size,
$\bar{f}^{1/3}$ and the distances are given in the units of the
current average diameter $d$ of the voids, calculated from their
volumes assuming that they are spherical (see text). The 
horizontal line is at ILD=0.5 diameters, the value we identify 
as the ILD$_c$. 
(Panel (b) is after Ref.~\onlinecite{prl_trial}.)}
\label{fig5}
\end{figure}

\begin{table}
\caption{\label{tab2} Critical strain $\varepsilon_c$ and critical
linear mean void size $\bar{f}_c^{1/3}$ from Fig.~\ref{fig5},
calculated as when the inter-void ligament distance crosses the line
ILD=0.5$d$: $\bar{f}_c^{1/3}=\bar{f}^{1/3}$(ILD=0.5$d$),
$\varepsilon_c=\varepsilon$(ILD=0.5$d$).}
\begin{ruledtabular}
\begin{tabular}{lccc}
ILD$_0$ & $\dot{\varepsilon}$ (sec$^{-1}$) & $\bar{f}_c^{1/3}$ & $\varepsilon_c$ (\%)  \\ \hline
1.81 & $10^8$ & 0.153 & 2.46 \\
1.81 & $10^9$ & 0.150 & 3.48 \\
1.50 & $10^9$ & 0.132 & 3.22 \\
1.20 & $10^9$ & 0.121 & 3.06 \\
1.00 & $10^9$ & 0.106 & 2.74
\end{tabular}
\end{ruledtabular}
\end{table}

Figure~\ref{fig3} offered several visual indications of the
interaction between voids.  Clearly, the separation between the void
surfaces (the ILD) serves as something akin to a reaction coordinate
for the coalescence: the voids coalesce when it goes to zero.  Other
indications include the displacement of the center of a void as it
grows preferentially toward the neighboring void and the change in the
void growth rate as the voids interact.  In this section we now
quantify two of these effects, the evolution of the ILD and the void
center movement, in order to analyze the coalescence. In
Section~\ref{sec_shape} we study the void shape evolution, 
and the void growth rate is studied in Section~\ref{sec_volume}.

In Figs.~\ref{fig5}(a) and (b) the dynamic evolution of the ILD has
been plotted for strain-rates $\dot{\varepsilon}=10^8$/sec and
$10^9$/sec and for various initial closest surface-to-surface
distances between the voids ILD$_0$. (In Figs.~\ref{fig2}-\ref{fig4}
the data from the case ILD$_0 \simeq$ 1.81 were shown.) The dynamic ILD
has been derived as the separation distance of the two surface atoms
from the two voids that are closest to the line connecting the
original centers of the voids.  The raw data of the ILD have been
plotted in Fig.~\ref{fig5}(a) and in Fig.~\ref{fig5}(b) scaled data
are shown. For the horizontal axis of Fig.~\ref{fig5}(b), the linear mean void size
$\bar{f}^{1/3}$ has been used in order to collapse data from different
strain-rates $\dot{\varepsilon}=10^8$/sec and $10^9$/sec in the same
figure. For the vertical axis the inter-void ligament distance has
been divided by the current diameter of a void, $d$ (note, that it is
not the initial value $d_0 = 2r_0$). 
The diameter $d$ is calculated from the void volume assuming a
spherical shape (the formula is given below).
This scaling is motivated by the ansatz that the
void diameter sets the length scale for the system, and hence
the relevant distance between the voids is not the pure distance, 
but its ratio with the diameters of the voids.  The void diameter
affects both the elastic and plastic fields around the void.
Initially the void separation
distance decreases essentially smoothly until plasticity begins,
eventually reaching zero. A transition occurs when the ILD starts to
decrease noticeably faster than the free impingement line (the thin solid
lines calculated from two independent spherical voids) indicating void
interactions at the onset of coalescence. The thin solid lines are
derived by calculating from the same simulations the average sizes of
the two voids $\bar{V}_{void}$ and deriving the diameter as
$d=2 \left( \frac{3}{4\pi} \bar{V}_{void} \right)^{1/3}$. Then
the free impingement curve is given by
${\mathrm{ILD}}_{free}~=~({\mathrm{ILD}}_0+d_0)(1+t\dot{\varepsilon})-d$.  
The accelerated ILD decrease associated with transition to coalescence 
takes place
when the ILD reaches approximately one half: ILD$_c =0.5 \pm 0.1$
diameter or one radius, independently of ILD$_0$ or the strain-rate.
Also note that a curve derived from a single void growth is provided
to estimate the contribution of uncorrelated faceting effects (the
``one void'' curve at ILD$_0$=1.81), and these effects are seen to be
relatively small.  They are most noticeable near the start of void
growth where there is a relatively large upward fluctuation in the
``one void'' curve.  The critical ILD of one radius is much lower than
the Brown-Embury estimate, and it corresponds to a strain of 3.48\%
($\bar{f}^{1/3} \simeq 0.15$) for ILD$_0$=1.81 at
$\dot{\varepsilon}=10^9$/sec, close to frame (c) of Fig.~\ref{fig3}.
The values for critical strain and linear mean void size when
ILD=0.5$d$, derived from Fig.~\ref{fig5}(b), are tabulated for the
simulated systems in Table~\ref{tab2}. In the very final stages the
ligament is drawn under biaxial stress, and the flow 
switches from radial material transport to tangential transport as the
mechanism switches from loop punching to drawing. This transition
is visible in
Figs.~\ref{fig5}(a) and (b) as slowing down of the reduction of
ILD. At this point, the material is highly defective but it remains
ductile.  There is no abrupt fracture, as might be expected at larger
length scales.  Here the final coalescence involves an extended
drawing and thinning of the ligament until rupture.

\begin{figure}
\includegraphics[height=8cm,angle=-90]{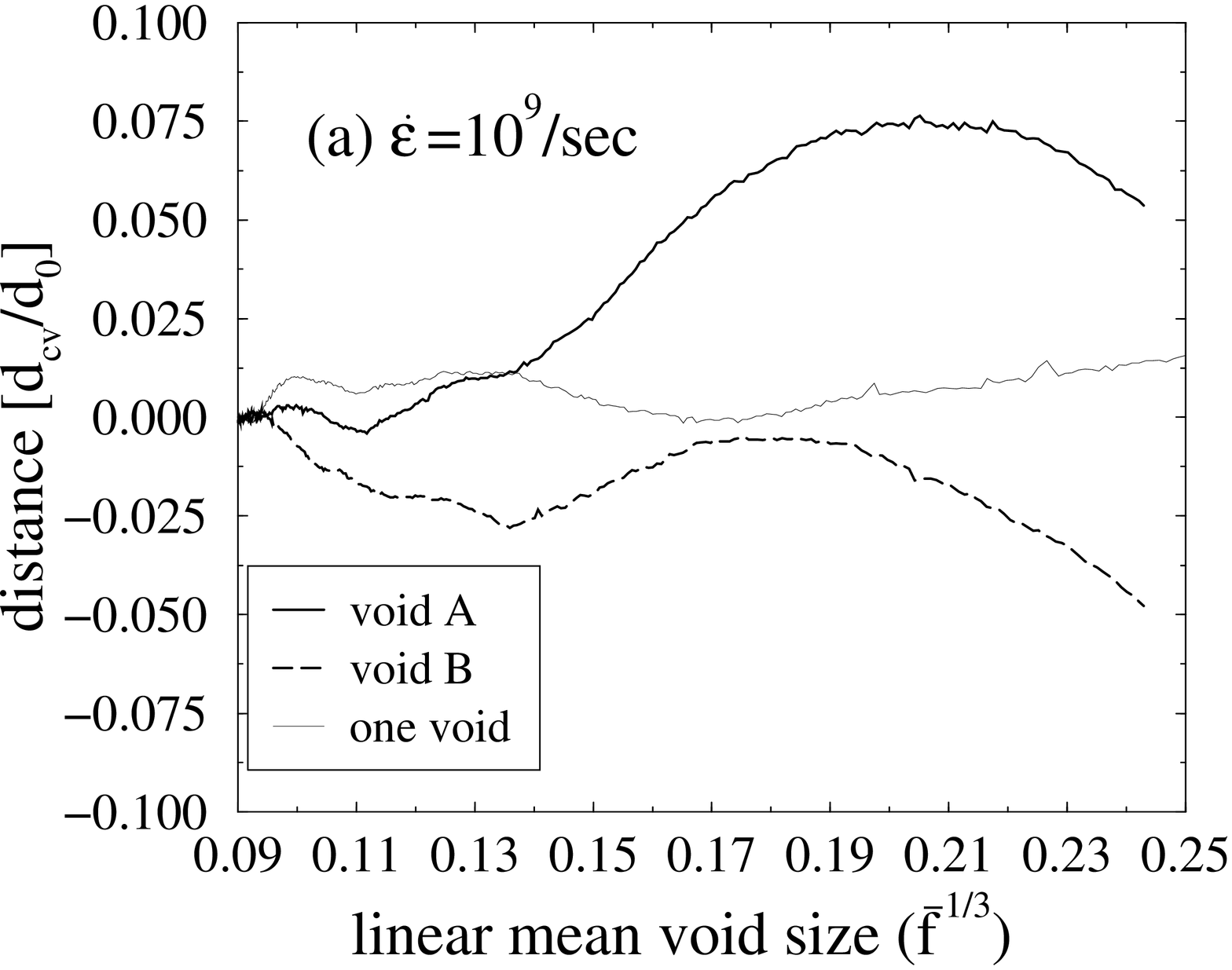}
\includegraphics[height=8cm,angle=-90]{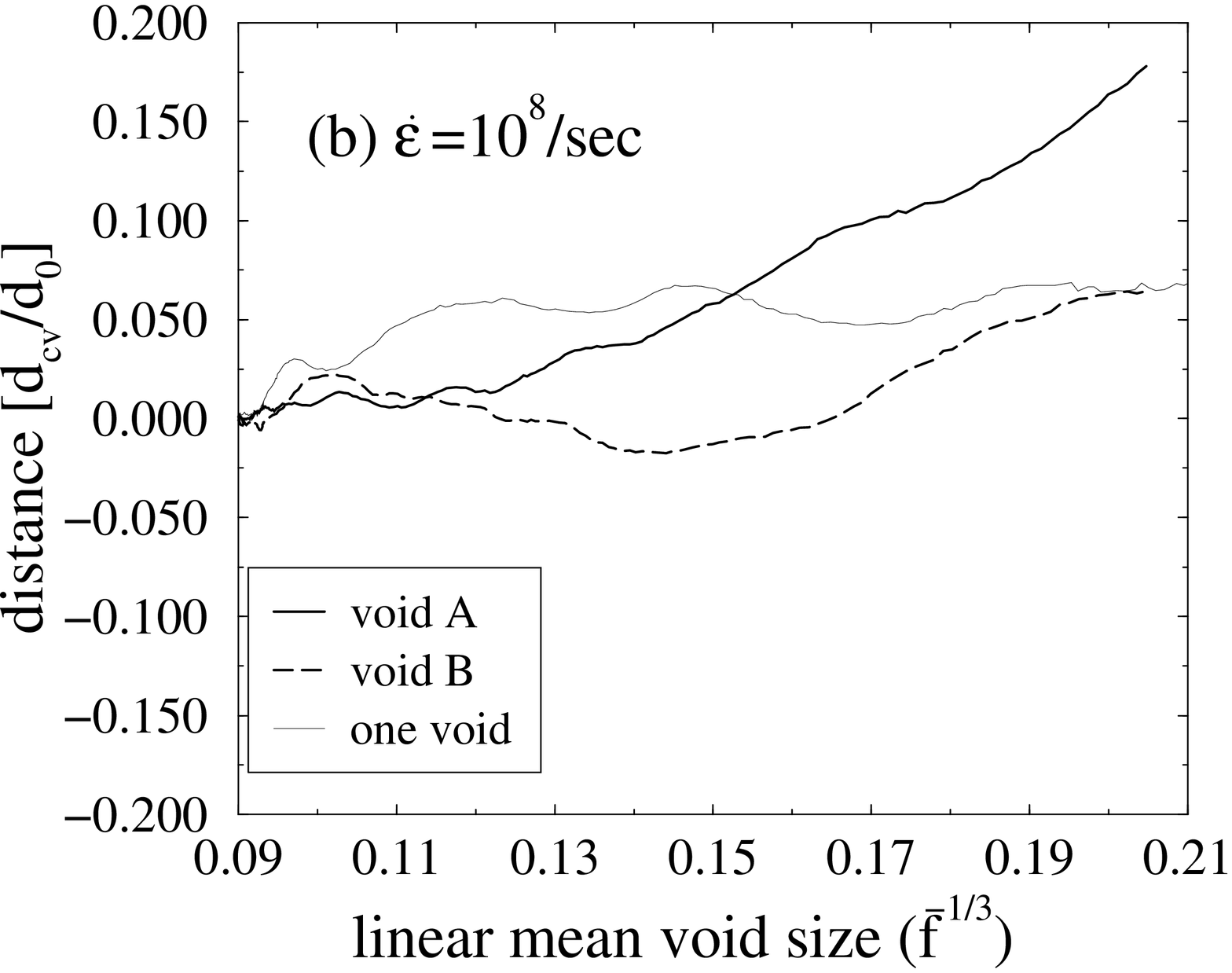}
\caption{Void center-of-mass displacement.  The distance $d_{cv}$ from 
the original center of void to the
instantaneous void center, projected onto the line connecting the
original void centers, is plotted versus the average void size to the
point of coalescence (ILD $\simeq 0$).  The sign of the displacement
$d_{cv}$ is positive for movement toward the other void.  Solid and long
dashed lines are for voids A and B, respectively. Strain-rates are
$\dot{\varepsilon}=10^9$/sec in (a) and $\dot{\varepsilon}=10^8$/sec
in (b).  The thin solid line is for a single void in the same size of
the box and with the same radius and strain-rate projected to the same
line. Here the distance $d_{cv}$ is given in the units of the original
void diameter $d_0$. ILD$_0$=1.81 in both (a) and (b).  
}
\label{fig6}
\end{figure}

Another measure of void interactions is whether the voids grow
preferentially toward their neighbor.  This effect is quantified in
Fig.~\ref{fig6}, which shows the movement of the center of mass of the
void surface for the voids shown in Fig.~\ref{fig3} (ILD$_0$=1.81).
We use here the center of mass of the void surface as the definition
for calculations of the void center. The void surface has been derived
by Voronoi triangulation based on the center points of the surface
atoms (the same is done when the volume of the void is
calculated). See Ref.~\onlinecite{prb} for the details of the void
surface derivation. The distance
\begin{equation}
d_{cv} =\left[\left(x'-x^0_{cv}\right)^2+\left(y'-y^0_{cv}\right)^2
+\left(z'-z^0_{cv}\right)^2\right]^{1/2}
\label{eq_dcv}
\end{equation}
is calculated between the original center of mass of the void surface
$[x^0_{cv},y^0_{cv},z^0_{cv}]$ and the projection $[x',y',z']$ of the
current center of void $[x_{cv},y_{cv},z_{cv}]$, where the projection 
is onto
the line connecting the original void centers. The sign is positive if
the void center has moved toward the other void and negative in the
opposite case.  

Let us look first at the case with
$\dot{\varepsilon}=10^9$/sec, Fig.~\ref{fig6}(a).  After the void
growth starts, the center of void A initially moves only slightly, but
at about $\bar{f}^{1/3} = 0.15$ (ILD =0.5$d$ in Fig.~\ref{fig5}(b) and
$\bar{f}_c^{1/3}$ in Table~\ref{tab2}), it starts to move in the
direction of the other void as the void growth becomes biased toward
its neighbor.  Just before coalescence the center of void A begins to
move away from void B, as the growth is biased in the opposite
direction. During this sequence, void B initially grows away from void
A, then roughly in unison with void A ($\bar{f}^{1/3} =
\bar{f}_c^{1/3} = 0.15$) it begins to grow toward its neighbor, and
before coalescence it too switches to growth away from the proximal
void. This retrograde growth happens at the same point (after
$\bar{f}^{1/3}=0.19$) as the decrease of the ILD begins to slow down in
Fig.~\ref{fig5}(b) [see also the snapshot in Fig.~\ref{fig3}(d)].  The
same phenomenon--first slow movement or repulsion from the void; then
growth toward the nearby void at about $\bar{f}^{1/3} = 0.15$
($\bar{f}_c^{1/3}$)--holds in the
$\dot{\varepsilon}=10^8$/sec case, too, Fig.~\ref{fig6}(b). However,
the retrograde growth phenomenon is less pronounced in the
$\dot{\varepsilon}=10^8$/sec case, as is the slowing down of the
decrease of the ILD in Fig.~\ref{fig5}(b). Indeed, only the growth
toward the neighboring void is well above the noise in the
$\dot{\varepsilon}=10^8$/sec case.  As a reference the movement of the
center of a single void (in same box size) projected to the same line
is plotted for both strain-rates, too.  Comparing the single
void case with the interacting voids with the same strain-rates, one
sees that the maximum distance the centers of the interacting voids
have moved is 2.5 to five times larger than the nanoscale random walk 
of the single void center, except for the void B in Fig.~\ref{fig6}(b).
The noise is these curves appears to be dependent on the history, 
reflecting the nature of the plastic deformation processes involved.  It 
is difficult to quantify the level of noise in such a non-Markov process,
but it should be clear that the
movement of the void center, especially the movement toward the
neighboring void, is statistically significant and not just due to 
fluctuations at the void surface.

\subsection{Shape Evolution of the  Voids}
\label{sec_shape}

\begin{figure}
\includegraphics[height=8cm,angle=-90]{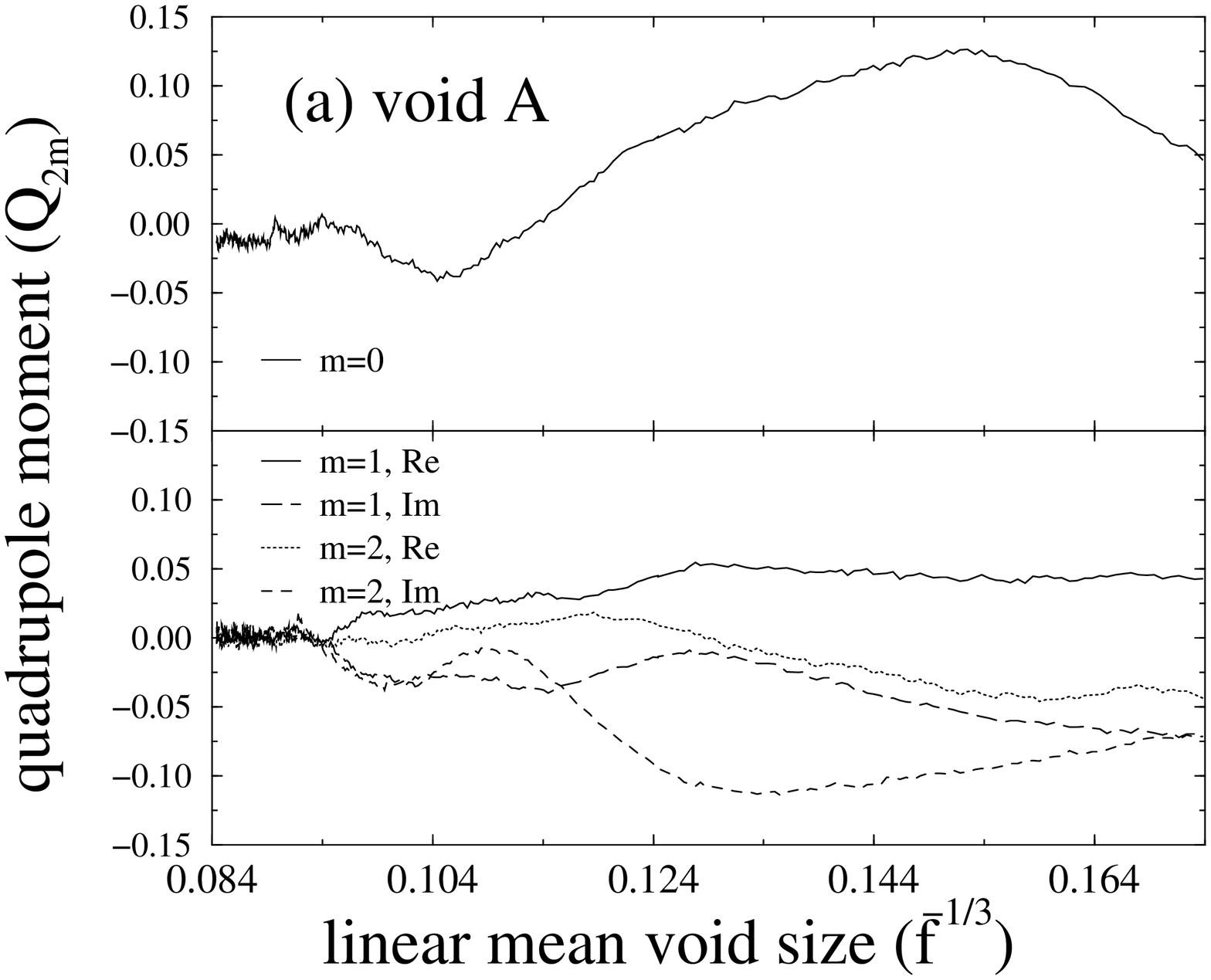}
\includegraphics[height=8cm,angle=-90]{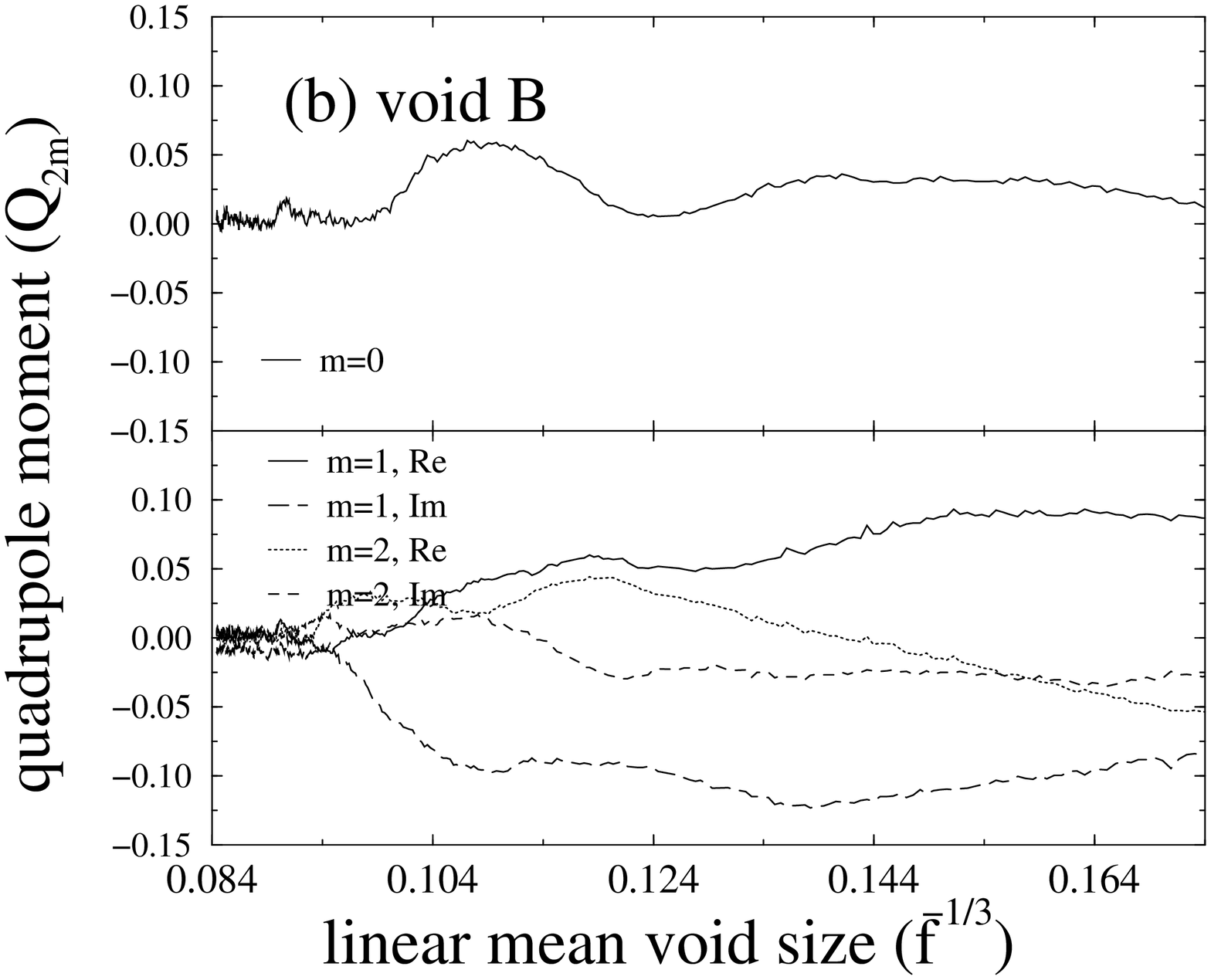}
\includegraphics[height=8cm,angle=-90]{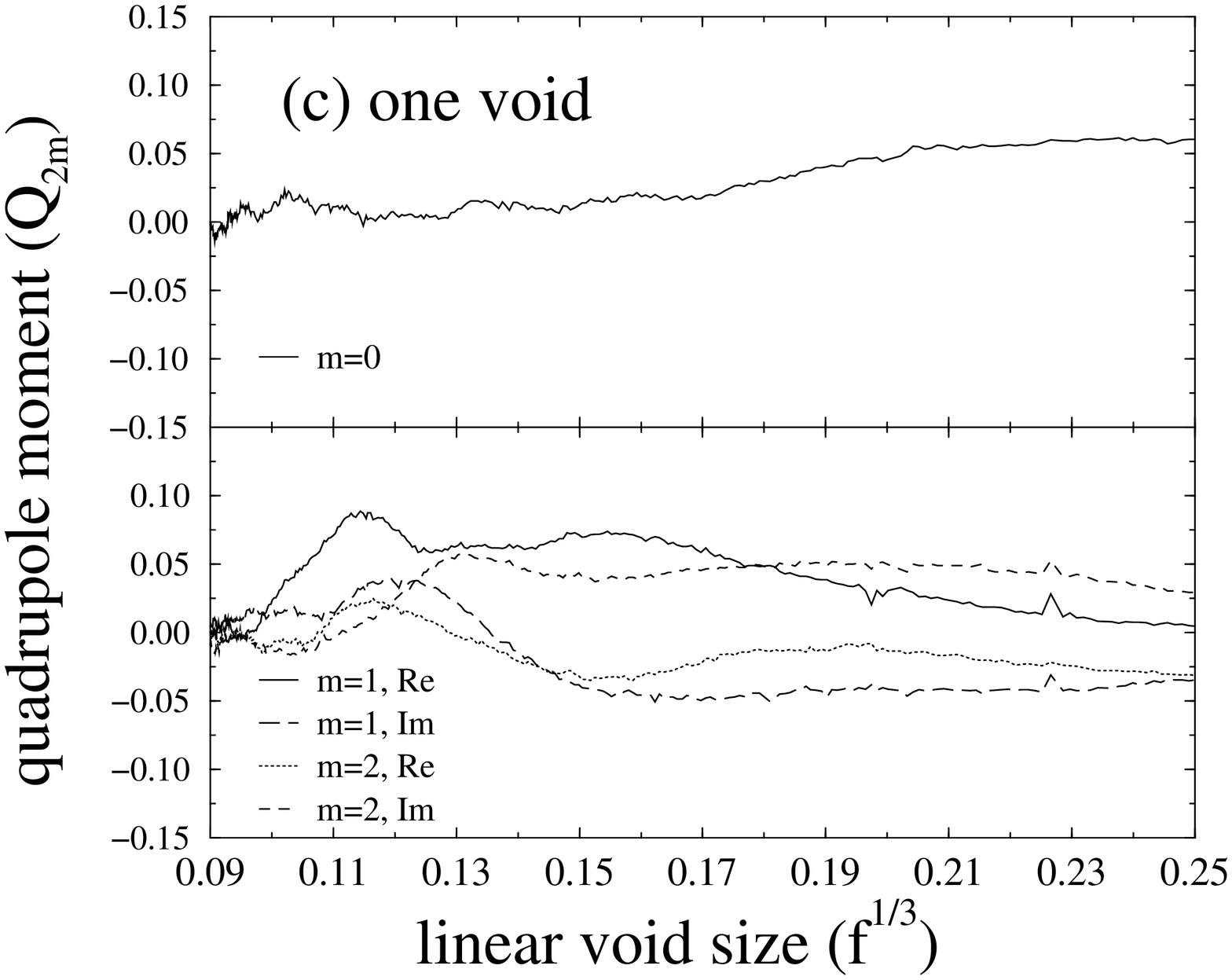}
\caption{(a) Quadrupole moments (\ref{eq_ql2}) of the surface of void A 
from the start of the simulation until coalescence.
The coordinate axes have been
rotated using Eq.~(\ref{eq_d}) so that the $z$-axis is aligned 
with the center of void B. 
The initial inter-void ligament distance is ILD$_0$=1.00 and the
strain-rate is $\dot{\varepsilon}=10^9$/sec. (b) Quadrupole moments for
the void B, located at $[0.1778, 0.0829, 0.0387]L$ from void A.  Now
Eq.~(\ref{eq_d}) has been used so that the positive $z$-axis is toward
the void A. 
(c) Quadrupole moments (not
rotated) for a
single void in a simulation with same box size as in (a) and (b) and
the strain-rate $\dot{\varepsilon}=10^9$/sec. }
\label{fig7}
\end{figure}

The presence of a nearby void not only affects the position of
the void but its shape, as well. The shape can be quantified by
calculating multipole moments of spherical harmonics $Q_{lm} \equiv
\frac{1}{\bar{r}^2} \, \int \!  Y_{lm}(\theta,\phi) \,
r^2(\theta,\phi) \, d\Omega$, see Refs.~\onlinecite{prb} and 
\onlinecite{stein}. 
Under fully triaxial expansion
the void tends to an octahedral shape because of the index of the 
active glide planes in FCC crystals, the anisotropic elastic constants, 
and anisotropic surface energies.  See also 
Ref.~\onlinecite{hori} for a study of void
shapes in FCC crystals.  On the other hand, under uniaxial expansion
the voids are of predominantly ellipsoidal shapes aligned along 
the preferred axis, and they make a transition from a prolate to an oblate
shape.~\cite{prb} 
References~\onlinecite{budiansky,Andersson,castaneda} have also
considered oblate void shapes under uniaxial loading through
continuum modeling. One may enquire whether the presence of
a nearby void causes evolution to an ellipsoidal shape,
see {\it e.g.} void A in Fig.~\ref{fig3}(c). Ellipsoidal
shapes can be quantified with quadrupole moments:
\begin{equation}\begin{array}{rcl}
Q_{20} & = & \frac{1}{4} \sqrt{ \frac{5}{\pi}} 
\frac{1}{\bar{r}^2} \int  \! 3 z^2 - r^2 \, d\Omega, \\
\mbox{Re } Q_{21} & =  & -\frac{1}{2} \sqrt{ \frac{15}{2\pi}} 
\frac{1}{\bar{r}^2} \int \!  xz \, d\Omega, \\
\mbox{Im } Q_{21} & = & -\frac{1}{2} \sqrt{ \frac{15}{2\pi}} 
\frac{1}{\bar{r}^2} \int \! yz \, d\Omega,\\
\mbox{Re } Q_{22} & = & \frac{1}{4} \sqrt{ \frac{15}{2\pi}} 
\frac{1}{\bar{r}^2} \int \! x^2 - y^2 \, d\Omega,\\
\mbox{Im } Q_{22} & = & \frac{1}{2} \sqrt{ \frac{15}{2\pi}} 
\frac{1}{\bar{r}^2} \int \! xy \, d\Omega,
\label{eq_ql2}
\end{array}
\end{equation}
where $\bar{r}^2 = \frac{1} {4 \pi} \int r^2(\theta,\phi) d\Omega$ is
averaged over the surface of the void. In calculating the quadrupole
moments the origin of the coordinates is taken to be the center of 
the void.  In Eq.~(\ref{eq_ql2}) the quadrupole moments are 
calculated with $\hat{z}$ as the preferred axis, whereas the
physically preferred axis is the line connecting the void centers.
Thus, we transform the moments to the more natural coordinates
using $D$-matrices~\cite{Rose}:
\begin{equation}
\begin{array}{rcl}
Q_{lm}(\theta', \phi') & = & \sum_{m'=-l}^l D^l_{m'm}(\alpha, \beta)
Q_{lm'}(\theta, \phi), \\
D^l_{m'm}(\alpha, \beta) & = & e^{-im'\alpha} d^l_{m'm}(\beta),
\label{eq_d}
\end{array}
\end{equation}
where the Euler angle $\alpha$ defines the rotation between
coordinates axes in $(xy)$ plane (corresponding angle $\phi$); and the
Euler angle $\beta$ describes the rotation in $z$-axis (corresponding
angle $\theta$). $d^l_{mm'}(\beta)$ can be found from
tables.~\cite{Hagiwara} The quadrupole moments are plotted for the voids A
and B from the simulation with ILD$_0$=1.00 in Figs.~\ref{fig7}(a) and
(b), respectively.  One sees from Fig.~\ref{fig7}(a) that void A
becomes markedly elliptical
in the direction of the other void, as represented by $Q_{20}$, when
$\bar{f}^{1/3}\simeq\bar{f}_c^{1/3} \simeq 0.106$ (from
Table~\ref{tab2}). The quadrupole data from an identical simulation
but with only one void are plotted in Fig.~\ref{fig7}(c) as a control,
and a smaller variation in the quadrupole moment is observed simply
due to fluctuations in the atomistic growth.
Other cases with
larger ILD$_0$ look about the same as Figs.~\ref{fig7}(a) and (b)
although the trend and especially the transition point 
may not be as clear. This variation may be due to the elastic and plastic
interactions causing random shape evolution longer before
the critical ILD$_c$ (ILD=0.5$d$) is reached in cases when the voids
are initially separated further from each other (larger ILD$_0$).  See
Fig.~\ref{fig7}(c) for the single void case as an example of the
random shape evolution. There $\bar{f}^{1/3}=0.09$ is when the
dislocation driven void growth starts, see inset of
Fig.~\ref{fig4}(b). Thus, in larger ILD$_0$ cases the random shape
evolution suppresses the transition to the shape evolution due to the
other void.

\section{Void Volume Evolution under the Influence of the Second Void}
\label{sec_volume}

In Ref. \onlinecite{prl_trial}, we briefly discussed how void
growth after the onset of interaction but prior to coalescence
is different from void growth for an isolated void.  We address
this further here.
How does the correlated growth differ from the exponential growth of 
an isolated void?  \cite{belak,prb} 

\subsection{A Pair of Voids}
\label{sec_volume_twovoids}

\begin{figure}
\includegraphics[height=8cm,angle=-90]{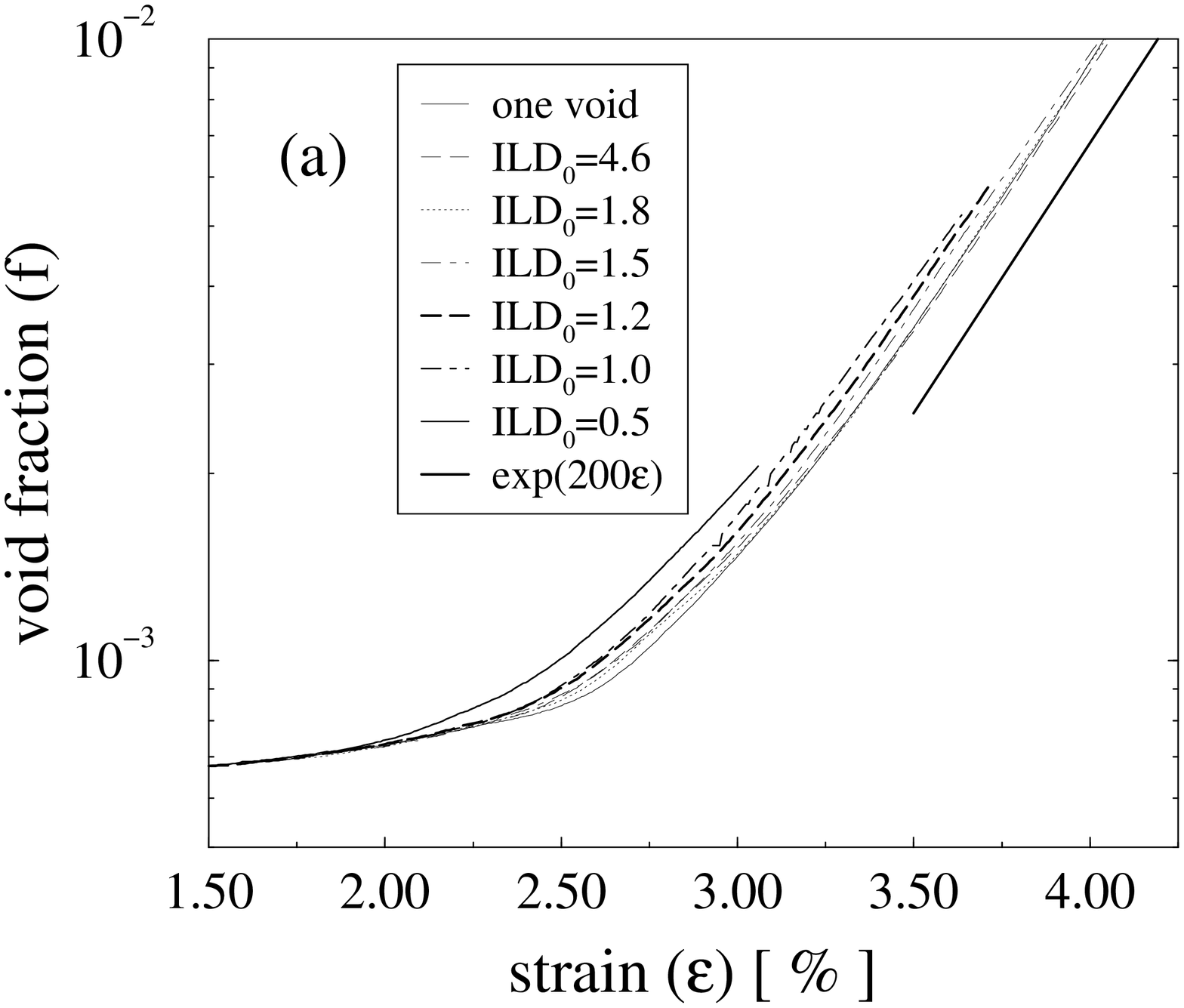}
\includegraphics[height=8cm,angle=-90]{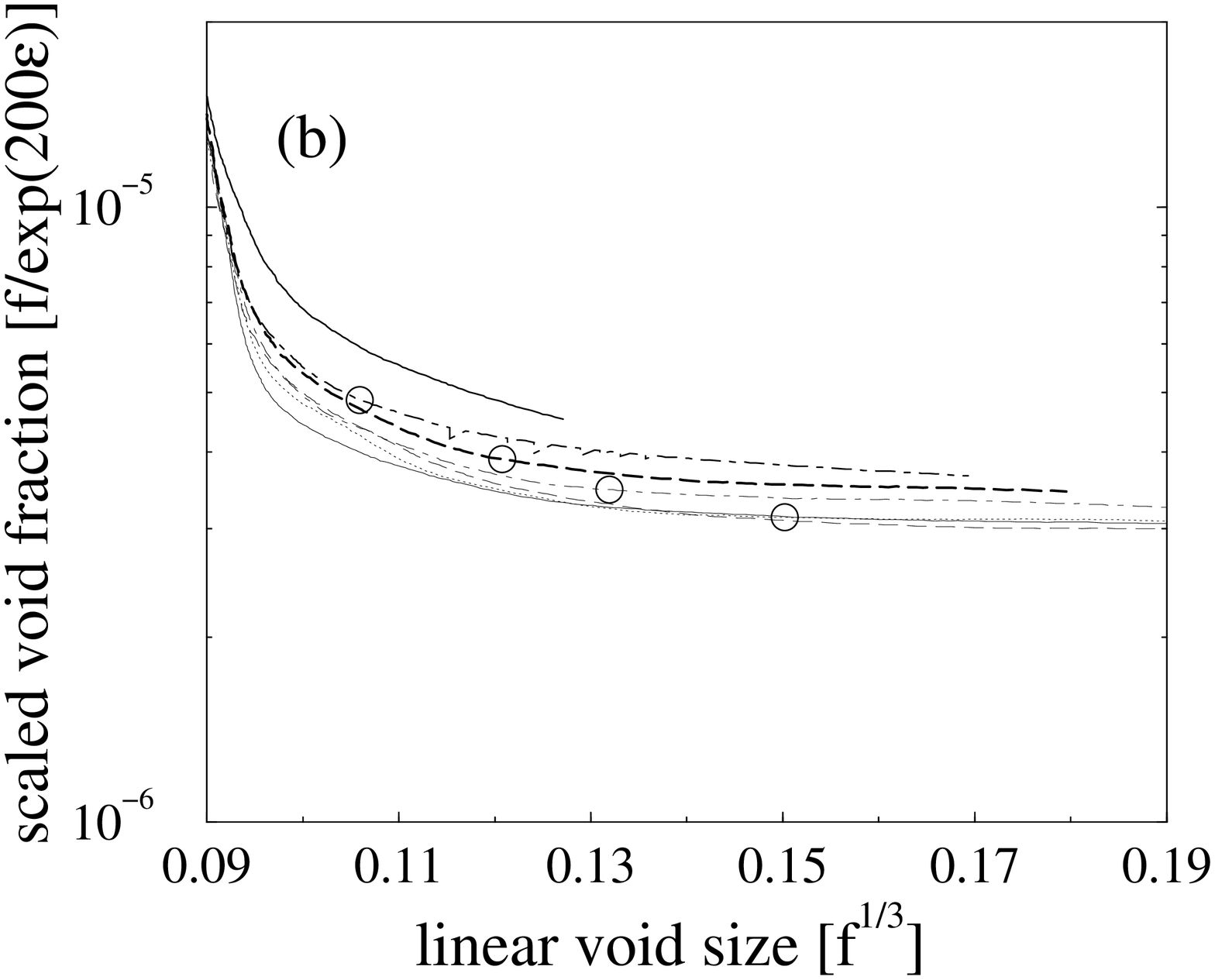}
\caption{(a) Growth of the void A until coalescence presented by void
fraction $f=V_{void}/V$ versus strain for ILD$_0$=0.50,
1.00,1.20,1.50, 1.81, and 4.62 diameters as well as in a single void
case (in the same box size) at $\dot{\varepsilon} =10^9$/sec.  
The asymptotic behavior (before finite size effects) is exponential
growth with $\exp(200\epsilon)$ as seen from the line drawn as a guide
to the eye. (b) The void size $f$ presented in (a) has been scaled
with the exponential and plotted versus linear void size
$f^{1/3}$. The circles point where the dynamical ILD's cross the
horizontal line ILD=0.5$d$ in Fig.~\ref{fig5}(b).}
\label{fig8}
\end{figure}

We first examine the volume evolution for the same two-void
simulations described above.  Figure~\ref{fig8}(a) shows the void
fraction, $f=V_{void}/V$, for the void A (also in the case of a single
void in the same box size) with respect to the strain before
coalescence. As can be seen from the figure the void grows as
$\exp(200\epsilon)$, at least for the larger ILD$_0$'s and for the
single void in the box. In Fig.~\ref{fig8}(b) we have factored out
the asymptotic growth rate from $f$ in order to emphasize the
differences between the curves and plotted versus linear void size
$f^{1/3}$.  The void growth data for ILD$_0$=4.62 and 1.81 coincide
with the single void curve.  The void growth rate with smaller
ILD$_0$'s reach their asymptotic growth rate earlier.  In the figure
we have drawn as circles the void size values, where the dynamic ILD's
cross the line ILD=0.5$d$ in Fig.~\ref{fig5}(b). As can be seen from figure,
there is no significant change in the void volume behavior when the voids
start to interact. Therefore we conclude that the void growth rate is
not affected by the interaction between the voids: thus the
interaction cannot be detected through the growth rate. 
The key factor for the void
growth rate is the rate at which the dislocations separate from the
void, and it appears to remain unchanged in the vicinity of the second
void. However, the location at the void surface from where the
dislocation loops separate is affected by the interaction with another
void, as seen in Section~\ref{sec_interaction} in the accelerated
reduction of ILD, in the movement of void center and in the shapes of
the voids.

\subsection{A Single Void Interacting with its Periodic Images}
\label{sec_volume_onevoids}

\begin{table}
\caption{\label{tab3} Sizes for simulated systems with various $r_0/L$
($r_0 = $ 2.17 nm is fixed) as the number of FCC cells, the
equilibrium side length of the cube $L$ at ambient pressure and room
temperature, and the number of atoms in the box after the void is
formed. The shortest initial inter-void ligament distance ILD$_0$ of the
void with its periodic image in the units of the void diameter $d$ is
reported in the last column.}
\begin{ruledtabular}
\begin{tabular}{ccccc}
$r_0/L$ & FCC cells & $L$ & atoms & ILD$_0$\\ \hline
1/3 & $18 \times 18 \times 18$ & 6.50 nm & 23328 & 0.50 \\
2/9 & $27 \times 27 \times 27$ & 9.75 nm & 78732 & 1.25 \\
1/6 & $36 \times 36 \times 36$ & 13.0 nm & 186624 & 2.00 \\
1/8 & $48 \times 48 \times 48$ & 17.3 nm & 442368 & 3.00 \\
1/10 & $60 \times 60 \times 60$ & 21.7 nm & 860396 & 4.50 \\
1/20 & $120 \times 120 \times 120$ & 43.3 nm & 6908379 & 9.00
\end{tabular}
\end{ruledtabular}
\end{table}

\begin{figure}
\includegraphics[height=8cm,angle=-90]{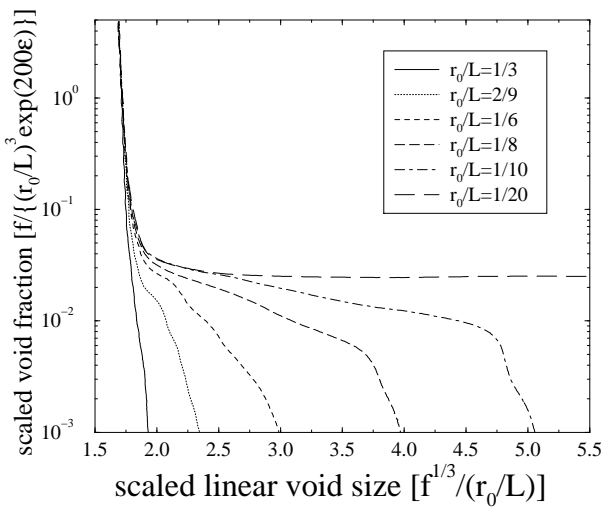}
\caption{Growth of a single void with varying initial box size. The
initial radius of the void is kept constant $r_0=2.17$nm, while the
initial side length of the cube is varied as $L=$6.50nm, 9.75nm,
13.0nm, 17.3nm, 21.7nm, and 43.3nm.  The void fraction $f$ has been
divided by $(r_0/L)^3$ in order to take into account different initial
volumes $V_0=L^3$ of the box. The vertical axis has been divided also
with the exponential of the strain in order to show the asymptotic
behavior as in Fig.~\ref{fig8} for the largest box size. The
strain-rate is for all the simulations $\dot{\varepsilon} =10^9$/sec.}
\label{fig9}
\end{figure}

We have also performed a series of simulations of a single void with
fixed initial radius size $r_0$ in various initial box sizes $V_0=L^3$ 
in order to find the coalescence process of the void with its (six)
periodic image(s), similar to the manner in which some continuum
calculations of coalescence have been done 
(cf.\ Refs.~\onlinecite{tvergaard} and \onlinecite{Andersson}). 
The details of the different box sizes are reported in Table~\ref{tab3}. 
Figure~\ref{fig9}
shows the data from this series of simulations. There we have scaled
the void fraction $f=V_{void}/V$ with the exponential,
$\exp(200\epsilon)$ (as in Fig.~\ref{fig8}) as well as $(r_0/L)^3$ in
order to take into account different initial volumes $V_0=L^3$ of the
box. Also $(r_0/L)$ scaling is applied to the linear void $f^{1/3}$
(horizontal axis). From the figure one concludes that the behavior is
opposite to the results from the case of an isolated pair of voids,
Section~\ref{sec_volume_twovoids}, in an important way.  The smaller
the box size, and hence the smaller the ILD$_0$, the later the void
starts to grow. This is easy to understand since measured from afar 
the two separate voids act as one big void, and thus grow faster than a
smaller void. In a single void with its periodic image that picture does 
not hold.  For example, there is no distance at which the stress field
approaches a single void stress field.  The periodic image only restricts 
the growth of the void.
At a more mechanistic level, the net effect of the array of voids
is to reduce the resolved shear stress driving dislocation emission
from the void surface, whereas a single nearby void enhances this
resolved shear stress on some regions of the void surface. An
intuitive way to understand this phenomenon is to consider that
the shear stress field of a single void forces interstitial loops
away from the void.  This field decreases with the distance
from the void as $1/r^3$.  At the near side of a neighboring void,
this field would tend to drive interstitial loops into that void;
in combination, it reduces that void's own stress field.  At the
far side, it adds to the other void's field, but the effect is
smaller due to the greater distance.  For a symmetric array, the
effect is to reduce the maximum resolved shear stress across the
surface and delay dislocation emission.

\ \\

\section{Conclusions}
\label{concl}

To summarize, interaction and coalescence of two voids in copper under
tension have been simulated in multi-million-atom MD simulations. The
effects of interactions between voids have been quantified by the
increased reduction-rate of their separation, by the movement of their
centers, and by their shape evolutions. The void interaction has also
been visualized by detecting the dislocations moving in the system
using the generalized centrosymmetry parameter.  The critical
inter-void ligament distance has been found to be close to one void
radius, independent of the strain-rate or the initial separation
distance ILD$_0$.  The onset of coalescence occurs at the point that
the plastic zones surrounding the voids first interact strongly. 
Signatures of coalescence have been found in the dynamic ILD curves
and the void center movements, as explained in detail here, including
reference to the stress-strain and void volume curves.  A weaker signature
of the onset of coalescence has also been found in the void shape 
curves giving the quadrupole moment evolution.
It has been demonstrated that the interaction of the voids is not
reflected in the volumetric asymptotic growth rate of the
voids. Finally, the coalescence process of an isolated pair of voids
has been shown to be markedly different than the coalescence of a
single void with its periodic images, so the latter would not provide
a reliable description of coalescence in typical low-symmetry
configurations.

In the future it would be interesting to study the cases with uniaxial
expansion (and the various orientations of the voids with respect to
the expansion direction), different sizes of the voids relative to
each other, other crystal structures as body-center cubic and
hexagonal lattices, and systems including larger collections of voids.

\begin{acknowledgments}

This work was performed under the auspices of the U.S.\ Dept.\ of
Energy by the University of California, Lawrence Livermore National
Laboratory, under contract no. W-7405-Eng-48.  We would like to
thank R.~Becker for useful discussions.

\end{acknowledgments}


\end{document}